\documentclass[12pt,a4paper]{article}
\usepackage{amsmath,amssymb}

\begin{document}

\title{Baryons in the nonperturbative string approach}

\author{I. M. Narodetskii and M. A. Trusov \\ \itshape ITEP, Moscow, Russia}

\maketitle

\begin{abstract}
\noindent We present some piloting calculations of masses and
short--range correlation coefficients for the ground states of
light and heavy baryons in the framework of the simple
approximation within the nonperturbative QCD approach.
\end{abstract}

The purpose of this talk is to discuss the results of the
calculation \cite{NT01} of the masses and wave functions of the
heavy baryons in a simple approximation within the nonperturbative
QCD (see \cite{Si99} and references therein). The starting point
of the approach is the Feynman--Schwinger representation for the
three quark  Green function in QCD in which the role of the time
parameter along the trajectory of each quark is played by the
Fock--Schwinger proper time. The proper and real times for each
quark related via a new quantity that eventually plays the role of
the dynamical  quark mass. The final result is the derivation
\cite{Si99} of the Effective Hamiltonian (EH). For the ground
state baryons without radial and orbital excitations in which case
tensor and spin-orbit forces do not contribute perturbatively the
EH has the following form
\begin{equation}
\label{EH} H=\sum\limits_{i=1}^3\left(\frac{{m_i^{(0)}}^2}{2m_i}+
\frac{m_i}{2}\right)+H_0+V,
\end{equation}
where $H_0$ is the non-relativistic kinetic energy operator and
$V$ is the sum of the perturbative one gluon exchange potential
$V_c$ and the string potential $V_{\mathrm string}$. The latter
has been calculated in \cite{FS91} as the static energy of the
three heavy quarks:
$V_{\mathrm{string}}(\mathbf{r}_1,\mathbf{r}_2,
\mathbf{r}_3)=\sigma R_{\mathrm{min}}$, where $R_{\mathrm{min}}$
is the sum of the three distances $|\mathbf{r}_i|$ from the string
junction point, which for simplicity is chosen as coinciding with
the center--of--mass coordinate.

In Eq. (\ref{EH}) $m_i^{(0)}$ are the current quark masses and
$m_i$ are the dynamical quark masses. In contrast to the standard
approach of the constituent quark model the dynamical mass $m_i$
is not  a free parameter  but it is expressed in terms of the
current mass $m^{(0)}_i$ defined at the appropriate scale of
$\mu\sim 1\mathrm{~GeV}$ from the condition of the minimum of the
baryon mass $M_B$ as function of $m_i$: $\partial
M_B(m_i)/\partial m_i=0 $. Technically, this has been done using
the einbein (auxiliary fields) approach, which is proven to be
rather accurate in various calculations for relativistic systems.

The EH  has been already applied to study baryon Regge
trajectories \cite{FS91} and very recently for computation of
magnetic moments of light baryons \cite{KS00}. The essential point
of this talk is that it is very reasonable that the same method
should also hold for hadrons containing heavy quarks. In what
follows we will concentrate on the masses of double heavy baryons.
As in \cite{KS00} we take as the universal parameter the QCD
string tension $\sigma$ fixed in experiment by the meson and
baryon Regge slopes. We also include the perturbative Coulomb
interaction with the frozen coupling
$\alpha_s(1\mathrm{~GeV})=0.4$.

We use the hyper radial approximation (HRA) in the hyper-spherical
formalism approach. In the HRA the three quark wave function
depends only on the hyper-radius
$R^2=\boldsymbol{\rho}^2+\boldsymbol{\lambda}^2$, where
$\boldsymbol{\rho}$ and $\boldsymbol{\lambda}$ are the appropriate
three-body Jacobi variables. Introducing the reduced function
$\chi(R)=R^{5/2}\psi(R)$ and averaging $V=V_c+V_{\mathrm{string}}$
over the six-dimensional sphere one obtains the Schr\"odinger
equation
\begin{equation} \label{shr}
\frac{d^2\chi(R)}{dR^2}+2\mu\left[E_n+\frac{a}{R}-bR-\frac{15}{8\mu
R^2}\right]\chi(R)=0, \end{equation} where
\begin{equation} \label{ab} a=\frac{2\alpha_s}{3}\cdot
\frac{16}{3\pi}\sum\limits_{i<j}\sqrt{\frac{\mu_{ij}}{\mu}},~~~
b=\sigma\cdot\frac{32}{15\pi}\sum\limits_{i<j}\sqrt{\frac{\mu(m_i+m_j)}{m_k(m_1+m_2+m_3)}},
\end{equation}
$\mu_{ij}$ is the reduced mass of quarks $i$ and $j$ and $\mu$ is
an arbitrary parameter with the dimension of mass which drops off
in the final expressions. We use the same parameters as in Ref.
\cite{KN00}: $\sigma=0.17\mathrm{~GeV}$, $\alpha_s=0.4$,
$m^{(0)}_q=0.009\mathrm{~GeV}$, $m^{(0)}_s=0.17\mathrm{~GeV}$,
$m^{(0)}_c=1.4\mathrm{~GeV}$, and $m^{(0)}_b=4.8\mathrm{~GeV}$.

The dynamical masses $m_i$ and the ground state eigenvalues $E_0$
calculated using the described above procedure are given for
various baryons in Table 1 of Ref. \cite{NT01}. For the light
baryons the values of light quark masses $m_q\sim
450-500\mathrm{~MeV}$ ($q=u,d,s$) qualitatively agree with the
results of Ref. \cite{KN00} obtained from the analysis of the
heavy--light ground meson states, but $\sim 60\mathrm{~MeV}$
higher than those of Refs. \cite{FS91}, \cite{KS00}. This
difference is due to the different treatment of the Coulomb and
spin--spin interactions. The light quark masses are increased by
$100-150\mathrm{~MeV}$ when going from the light to heavy baryons.
For the heavy quarks ($c$ and $b$) the variation in the values of
their dynamical masses in different baryons is marginal. Note that
the masses of the light quarks in baryons are slightly smaller
than those in the mesons.

For many applications the quantities $ R_{ijk}=\langle\psi_{ijk}|
\delta^{(3)}(\mathbf{r}_j-\mathbf{r}_i)|\psi_{ijk}\rangle$ are
needed. Note that these quantities depend on the third or
`spectator' quark through the three--quark wave function. To
estimate effects related to the baryon wave function we solve Eq.
(\ref{shr}) by the variational method using a simple trial
function $\chi(R)\sim R^{5/2}e^{-\mu \beta^2R^2}$, where $\beta$
is the variational parameter. Then
$R_{ijk}=\left(2\beta^2\mu_{ij}/\pi\right)^{3/2}$. The results of
the variational calculations are given in Table 3 of \cite{NT01}.
Comparing the results with those of Ref. \cite{KN00} we confirm
the inequalities $R_{ijk}<\frac{1}{2}R_{ij}$ and
$R_{ijk}>R_{ijl}$, if $m_k\le m_l$, first suggested in Ref.
\cite{Li96} from the observed mass splitting in mesons and
baryons. Here $R_{ij}$ is the corresponding quantity for a meson.
In particular, we obtain $R_{ijk}/R_{ij}=0.44$, $0.40$, $0.37$,
and $0.34$ for $ijk=ucd$, $scu$, $ubd$, and $sbu$, respectively.
These estimations agree with the results obtained using the
non--relativistic quark model or the bag model or QCD sum rules
which are typically in the range $0.1-0.5$. Note also that if
$i,j$ are the light quarks, and the quarks $k$ and $l$ are the
heavy then $R_{ijk}\approx R_{ijl}$ ({\it i.g.} $R_{qqc}\approx
R_{qqb}$) in agreement with the limit of the heavy quark effective
theory.

Note also that the wave function calculated in HRA show the
marginal diquark clustering even in the doubly heavy baryons .
E.g. in the $qcc$ baryon $\bar{r}_{qc}=0.61\text{~fm}$ while
$\bar{r}_{cc}=0.45\text{~fm}$. Likewise
$\bar{r}_{qb}=0.53\text{~fm}$ and $\bar{r}_{bb}=0.25\text{~fm}$ in
the $qbb$ baryon. This is principally kinematic effect related to
the fact that in the HRA the difference between the various
$\bar{r}_{ij}$ in a baryon is due to the factor
$\sqrt{1/\mu_{ij}}$ which varies between $\sqrt{2/m_i}$ for
$m_i=m_j$ and $\sqrt{1/m_i}$ for $m_i\ll m_j$. For the light
baryons $\bar{r}_{qq}\sim 0.7-0.8\text{~fm}$.

To calculate hadron masses we, as in Ref. \cite{FS91}, first
renormalize the string potential:  $V_{\mathrm{string}}\to
V_{\mathrm{string}}+\sum\limits_iC_i$, where the constants $C_i$
take into account the residual self-energy (RSE) of quarks. In the
present work we treat them phenomenologically. We adjust $C_i$ to
reproduce the center-of-gravity for baryons with a given flavor.
To this end we consider the spin-averaged masses, such as:
$(M_N+M_{\Delta})/2$, and
$(M_{\Lambda}+M_{\Sigma}+2M_{\Sigma^*})/4$ and analogous
combinations for $qqc$ and $qqb$ states. Then we obtain
$C_q=0.34$, $C_s=0.19$, $C_c\sim C_b\sim 0$.

We keep these parameters fixed to calculate the masses given in
Table 1, namely the spin--averaged masses (computed without the
spin--spin term) of the lowest double heavy baryons. In this Table
we also compare our predictions with the results obtained using
the additive non--relativistic quark model with the power-law
potential \cite{BDGNR94}, relativistic quasipotential quark model
\cite{E97}, the Feynman--Hellmann theorem \cite{LRP95} and with
the predictions obtained in the  approximation of double heavy
diquark \cite{LO99}. \vspace{0.5cm}

\centerline{\small {\bfseries Table 1} Masses of baryons
containing two heavy quarks.}

\begin{center}
\begin{tabular}{|c|c|c|c|c|c|}
\hline State & present work & Ref. \cite{BDGNR94}& Ref.
\cite{E97}& Ref. \cite{LRP95}& Ref. \cite{LO99}\\ \hline
$\Xi\{qcc\}$    & 3.69 & 3.70 & 3.71 & 3.66 & 3.48\\
$\Omega\{scc\}$ & 3.86 & 3.80 & 3.76 & 3.74 & 3.58\\ \hline
$\Xi\{qcb\}$    & 6.96 & 6.99 & 6.95 & 7.04 & 6.82      \\
$\Omega\{scb\}$  & 7.13 & 7.07 & 7.05 & 7.09 & 6.92     \\ \hline
$\Xi\{qbb\}$    & 10.16 & 10.24 & 10.23 & 10.24 & 10.09
\\ $\Omega\{sbb\}$   & 10.34 & 10.30 & 10.32 & 10.37 & 10.19
\\ \hline
\end{tabular}\end{center}

In conclusion, we have employed the general formalism for the
baryons, which is based on nonperturbative QCD and where the only
inputs are the string tension $\sigma$, the strong coupling
constant $\alpha_s$ and two additive constants, $C_q$ and $C_s$,
the residual self--energies of the light quarks. Using this
formalism we have  performed the calculations of the
spin--averaged masses of baryons with two heavy quarks. One can
see from Table 1 that our predictions are especially close to
those obtained in Ref. \cite{BDGNR94} using a variant of the
power--law potential adjusted to fit ground state baryons.

One of the authors (I.M.N.) thank W.Briscoe and H.Haberzettl for
organizing an excellent Conference with a stimulating scientific
program. This work was supported in part by RFBR grants Refs.
00-02-16363 and 00-15-96786.

\end{document}